\documentstyle[12pt,epsfig]{article}

\newcommand{\be}{\begin{equation}}
\newcommand{\ee}{\end{equation}}

\def\aprge{\buildrel > \over {_{\sim}}}
\begin{document}  
\topmargin 0pt
\oddsidemargin=-0.4truecm
\evensidemargin=-0.4truecm
\renewcommand{\thefootnote}{\fnsymbol{footnote}}
\newpage
\setcounter{page}{1}
\begin{titlepage} 
\vspace*{-2.0cm}
\begin{flushright}
FISIST/17-99/CFIF \\
hep-ph/9910433
\end{flushright}
\vspace*{0.5cm}   
\begin{center}     
{\Large \bf Comment on ``New conditions for a total neutrino \\
\vspace{0.07cm}
conversion in a medium''}\\
\vspace{1.0cm}
{\large E. Kh. Akhmedov$^{1,2}$ 
\footnote{E-mail: akhmedov@gtae2.ist.utl.pt}
and A. Yu. Smirnov$^{3,4}$
\footnote{E-mail: smirnov@ictp.trieste.it}}\\
\vspace*{0.15cm}
$^1${\em Centro de F\'\i sica das Interac\c c\~oes Fundamentais (CFIF), 
Departamento de Fisica, Instituto Superior T\'ecnico, 
Av. Rovisco Pais, P-1049-001 Lisboa, Portugal}\\
\vspace{0.05cm}
$^2${\em National Research Centre Kurchatov Institute, Moscow 123182,
Russia}\\
\vspace{0.05cm}
$^3${\em International Centre for Theoretical Physics, Strada Costiera 
11, I-34100 Trieste, Italy}
\vspace{0.05cm}
$^4${\em Institute for Nuclear Research, RAS, Russia}
\end{center}
\vglue 0.8truecm
\begin{abstract}
We show that the conditions for total neutrino conversion 
found in \cite{CP1} are equivalent to the conditions of maximal depth 
(parametric resonance) and ($\pi/2+\pi k$) -  phase of parametric 
oscillations. Therefore the effects considered in \cite{CP1} are 
a particular case of the parametric resonance in neutrino oscillations.  
The existence of strong  enhancement peaks in transition probability $P$ 
rather than the condition $P=1$ is of physical relevance.  
We comment on possible realizations and implications of the parametric
enhancement of neutrino oscillations. 
\end{abstract}

\end{titlepage}
\renewcommand{\thefootnote}{\arabic{footnote}}
\setcounter{footnote}{0}
\newpage

1. In the present Comment we show that the conditions for total neutrino 
conversion studied by Chizhov and Petcov \cite{CP1} are just the conditions  
of the parametric resonance of neutrino oscillations supplemented by 
the requirement that the parametric enhancement be complete. Therefore 
the ``new effect of total neutrino conversion'' \cite{CP1} is nothing but
a particular case of the parametric enhancement of neutrino oscillations, 
suggested in \cite{ETC,Akh1} and widely discussed in the literature 
\cite{KS,LS,Akh2,ADLS,Akh3}. 

The parametric resonance occurs when the oscillation frequency changes in a 
certain correlation with the frequency itself and with the amplitude of the 
oscillations, leading to specific phase relationships. A classical example is 
a pendulum with vertically oscillating suspension point \cite{LL}. 
This situation can, in particular, be realized for oscillating neutrinos 
crossing layers of medium of different densities \cite{ETC,Akh1}. Indeed,
the oscillation parameters depend on matter density, and crossing the layers 
of different density means changing frequency and amplitude of neutrino 
oscillations.

2. The propagation of neutrinos through medium with periodic density
modulations leads to {\it parametric oscillations} \cite{ETC,Akh1}, see
figs. 1 and 2.  
Let us consider neutrino propagation in a medium with the periodic 
``castle wall" density profile: a system of alternating layers of matter 
with constant densities $N_1$ and $N_2$ and widths $L_1$ and $L_2$. 
Let $\theta_{1,2}$ be the mixing angles in matter at densities $N_1$ and 
$N_2$. We denote by $2\phi_i$ ($i=1,2$) the oscillation phase acquired by
neutrinos in the layer of density $N_i$ and width $L_i$. We will use the 
notation $s_i \equiv \sin\phi_i$, $c_i \equiv\cos \phi_i$. The evolution
matrix over one period of density modulation $L=L_1+L_2$ is \cite{Akh2} 
\be
U_2=Y-i\mbox{\boldmath $\sigma$} {\bf X}=
\exp[-i(\mbox{\boldmath $\sigma$}{\bf \hat{X}}) \Phi] \,, 
\label{UT}
\ee
\be
Y=c_1 c_2-\cos(2\theta_1-2\theta_2) s_1 s_2\,, \quad \; 
\Phi=\arccos Y=\arcsin |{\bf X}|\,,\quad\;{\bf \hat{X}}={\bf X}/|{\bf X}|\,. 
\label{Y}
\ee
The vector ${\bf X}$ can be written in
components as 
\be
{\bf X}=\left((s_1 c_2 \sin 2\theta_1+s_2 c_1 \sin 2\theta_2),\; 
-s_1 s_2 \sin (2\theta_1-2\theta_2),\; 
-(s_1 c_2 \cos 2\theta_1+s_2 c_1 \cos 2\theta_2 )\right).
\label{comp}
\ee
Notice that $Y^2+{\bf X}^2=1$ as a consequence of unitarity of $U_T$.

\begin{figure}[htb]
\hbox to \hsize{\hfil\epsfxsize=12cm\epsfbox{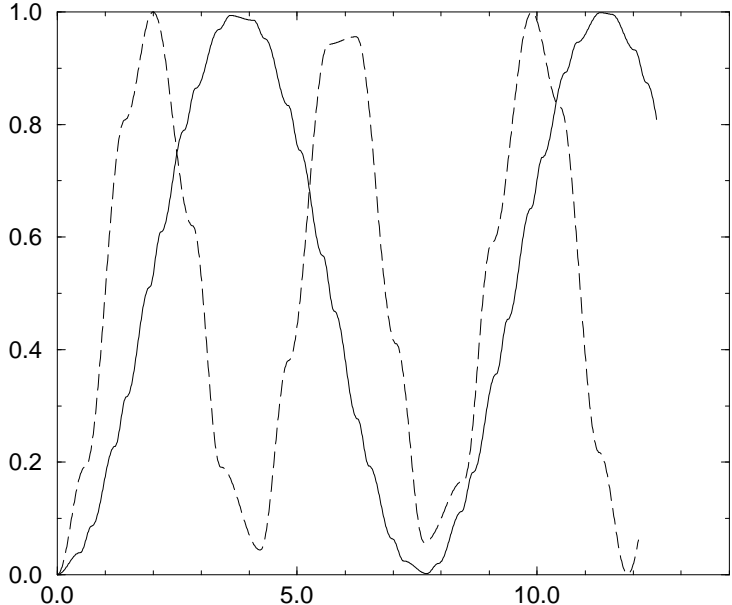}\hfil}
\caption{\small
Parametric oscillations in a medium with ``castle wall'' density
profile for the case $X_3=0$ (parametric resonance).
Solid curve: transition probability for neutrino flavour
oscillations as a function of the coordinate along the neutrino path
for the case of total conversion over 5 periods of density modulation
(10 layers). Dashed curve: the same for the case of total conversion
over 3 layers. The kinks correspond to the borders of the layers of
different densities. The curves were plotted for the realization (\ref{res2}) 
($c_1 = c_2 = 0$) of the parametric resonance condition. Neutrino energy is 
between the MSW resonance energies corresponding to the densities $N_1$ 
and $N_2$.} 
\label{param1}
\end{figure}

{}From Eq. (\ref{UT}) one easily finds the transition probability 
after passing $n$ periods \cite{Akh1,Akh2} 
\be
P(\nu_a\to \nu_b; r=nL) = \left(1 - \frac{X_3^2}{{\bf X}^2}\right)\sin^2
\Phi_p\,, \quad \quad \Phi_p =  n\Phi ~. 
\label{prob1}
\ee
The transition probability after passing an odd number of alternating 
layers, which can be considered as 
$n$ periods plus one additional layer of density $N_1$ 
(the corresponding distance $r=nL+L_1$), is also given by 
Eq. (\ref{prob1}), the only difference being that the phase is now 
\be
\Phi_p=n\Phi+\varphi\,,\quad\quad
\varphi = \arcsin\left(s_1 \sin 
2\theta_1/\sqrt{1-X_3^2/|{\bf X}|^2}\right)\,. 
\label{varphi}
\ee
Eqs. (\ref{prob1}) and (\ref{varphi}) give the transition  
probability at the borders of the layers. 
The pre-sine factor in (4) and $\Phi_p$ are the depth and the
phase of the parametric oscillations.  
The phase $\Phi_p$ determines the length of the parametric
oscillations (see figs. 1 and 2).

The parametric resonance occurs when the pre-sine factor in (\ref{prob1})
becomes equal to unity, {\it i.e.} the depth of the parametric
oscillations is maximal. 
The resonance condition is therefore (see Eq.~(26) in \cite{Akh2})  
\be
X_3 \equiv -(s_1 c_2 \cos 2\theta_1 +s_2 c_1 \cos 2\theta_2 )=0\,.
\label{res1}
\ee
The parametric resonance condition (\ref{res1}) can be realized in 
various ways 
\footnote{We do not consider the trivial cases of 
the MSW resonance for which $X_3 = 0$ because $\cos 2\theta_{i} = 0$  
and $s_{i} = \pm 1$, $i = 1$ or 2,  or $\cos 2\theta_1=\cos 2\theta_2=0$.}.  
One well known realization \cite{ETC,Akh1,KS,LS,Akh2} is 
$c_1 = c_2  = 0$, or 
\be
2\phi_1 = \pi + 2\pi k'\,, \quad\quad  2\phi_2 = \pi + 2\pi k^{''} \,, 
\label{res2}
\ee 
independently of the mixing angles \footnote 
{It was renamed into  ``the oscillation length resonance'' in \cite{PE}.}.  
(This was reproduced as solution III in \cite{CP1}, see Eq. (18) there).
If, however, $c_1$ and $c_2$ are non-zero, the cancellation between the two 
terms in  (\ref{res1}) can occur, which implies certain correlation between 
the phases and mixing angles in the layers. (This covers solution IV in
\cite{CP1}). For an example, see fig. 2. 

In general, the parametric resonance in neutrino oscillations 
does not require a periodic matter density profile (although the periodicity 
may make it easier to meet the resonance conditions), and can occur even in 
stochastic media \cite{KS}. 

\begin{figure}[htb] 
\hbox to \hsize{\hfil\epsfxsize=12cm\epsfbox{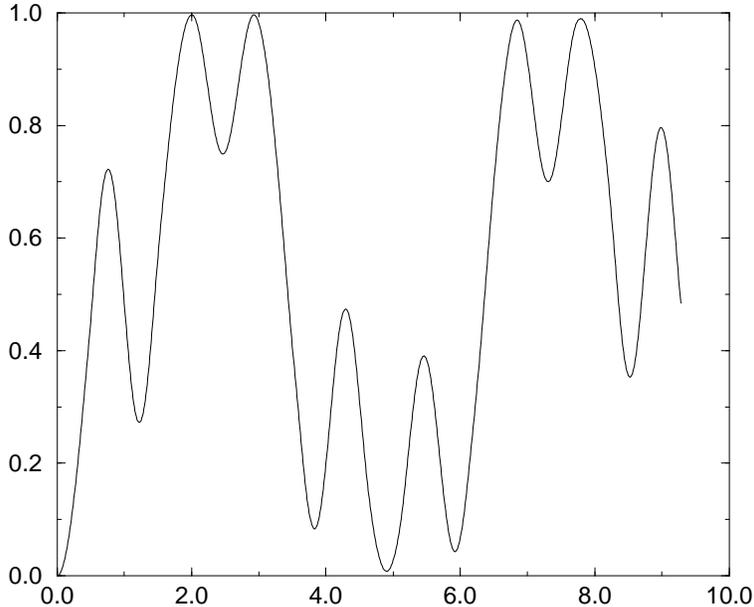}\hfil}
\caption{\small
Same as in fig. 1 but for the case when the parametric resonance condition 
is realized through the cancellation of the two terms in Eq. (\ref{res1}). 
Total conversion is achieved over 3 layers. (Similar dependence of the 
transition probability can be obtained in the case $c_1 = c_2 = 0$ provided 
that the neutrino energy is above the MSW resonance energies corresponding to 
the densities $N_1$ and $N_2$). }
\label{param2}
\end{figure}

3. As follows immediately from (\ref{prob1}), the conditions for 
total neutrino conversion $P(\nu_a\to \nu_b) = 1$ are
\be
X_3 = 0\,, \quad \quad \Phi_p = \frac{\pi}{2} + 2\pi k\, 
\label{cond}
\ee
for evolution over any number of layers, including the two- and 
three-layer cases considered in \cite{CP1}
\footnote{These cases correspond to $n=1$ in Eqs. (\ref{prob1}) and 
(\ref{varphi}).}.   
Thus {\it the maximal transition probability implies the fulfillment of
the parametric resonance condition.} 

According to  Eq. (\ref{prob1}) for two layers the conditions (\ref{cond}) 
reduce to   
\be 
X_3 = 0\,, \quad \quad Y = 0\,.  
\label{res3}
\ee
It is easy to see that Eqs. (\ref{res3}) are equivalent to two conditions 
(22) in \cite{CP1} (solution IV). 

Notice that conditions (\ref{res3}) can be obtained  directly from (\ref{UT}): 
the survival probability $P(\nu_a\to \nu_a) = Y^2 + X^2_3$ and therefore
the condition of total neutrino conversion gives  $Y^2 + X^2_3 = 0$.

Consider now the three-layer case (it gives a good approximation for 
the case of neutrinos crossing the earth, where the layers correspond 
to the mantle, core and then again mantle).  
Using expression for the phase (\ref{varphi}), one can write the conditions 
of total transition (\ref{cond}) as $X_3 = 0$, $Y = \pm s_1\sin2 \theta_1$, 
or equivalently as  
\be
X_3 = 0, \quad \quad  2 c_1 Y - c_2 = 0\, .   
\label{Z1}        
\ee
Conditions (\ref{Z1}) can also be obtained  directly  from the
evolution matrix which in this case is \cite{Akh2} 
\be
U_3 = Z-i\mbox{\boldmath $\sigma$}{\bf W}\,.
\label{U3}
\ee
Here 
\be
Z=2 c_1 Y-c_2 \,,  
\label{Z}
\ee
$Y$ has been defined in (\ref{Y}), and the vector ${\bf W}$ can be 
written in components as
\be
{\bf W}=\left( 2 s_1 Y \sin 2\theta_1+s_2 \sin 2\theta_2 \,, ~0\,,
~-\left(2 s_1 Y \cos 2\theta_1+ s_2 \cos 2\theta_2\right)
\right) \,.
\label{W}
\ee
The neutrino flavour transition probability in this case is 
$P(\nu_a\to \nu_b)=W_1^2$. The total neutrino conversion corresponds to 
zero survival probability: $Z^2 + W_3^2 = 0$, or 
\be
Z = 0\,, \quad \quad  W_3 = 0\,. 
\label{cond2}
\ee
There are two possible realizations of these conditions, depending on the
value of $c_1$. If $c_1 = 0$ then from (\ref{Z}) and (\ref{cond2}) it follows 
that $c_2$ must vanish, too. So, we arrive at the realization (\ref{res2}) of 
the parametric resonance condition (\ref{res1}). The second condition in 
(\ref{cond2}) is then the one for the ``totality'' of transition. It can be 
written as $\cos(2\theta_1 - 2\theta_2) = \cos 2\theta_2/2\cos 
2\theta_1$, which is equivalent to the requirement that the transition 
probability 
\cite{LS} 
\begin{equation}
P= \sin^2 (2\theta_2 - 4\theta_1)
\label{probmax}
\end{equation}
takes the value 1. 

If $c_1 \neq 0$ then the first equality in (\ref{cond2}) implies $Y=c_2/2c_1$. 
Inserting this into the expression for $W_3$ in (\ref{W}) one obtains 
$W_3=X_3/c_1$. 
The condition $W_3=0$ thus means $X_3=0$. Therefore in this case, too,  
total neutrino conversion implies parametric resonance. 
Conditions (\ref{Z1}) are equivalent at $c_1 \neq 0$ to the conditions of 
the total neutrino conversion in Eq. (26) of \cite{CP1}. 

Similarly, one can analyze the case of $\nu_2 \to \nu_e$ transitions which 
is relevant for oscillations of solar  and supernova  neutrinos in the
earth. In particular, it is easy to show that the parametric resonance 
condition for the probability $P_{2e}$ of $\nu_2\to \nu_e$ oscillations is 
\be 
X_3'\equiv X_3 \cos\theta_0 - X_1 \sin \theta_0 =0. 
\label{co2}
\ee 
The conditions of total $\nu_2\to \nu_e$ conversion found in [1] imply 
equality (\ref{co2}).

4. The existence of strong  enhancement peaks in transition probability $P$ 
rather than the condition $P=1$ is of physical relevance.  
For sufficiently large  vacuum mixing angles, 
the transition probability has a series of peaks of comparable 
height and the total conversion peak is just one of them. In fact, peaks
with $P_{max} <1$ can contribute to observable effects even more than the 
ones with $P_{max} = 1$. For some applications, {\it e.g.}, for oscillations 
of solar neutrinos in the earth, even partial (or relative) enhancement
can be important. 

Let us  comment on various realizations of the parametric enhancement of 
neutrino oscillations. Large oscillation effects can be due to large mixing 
in matter and therefore to large-amplitude oscillations, or due to specific 
properties of the density profile. In general, both mechanisms are present. 
Depending on neutrino parameters, either of the mechanisms can dominate, or 
they can give comparable contributions to the observable effects.  

(i) The most interesting case is the one when neutrino mixing in matter of 
both densities $N_1$ and $N_2$ is small: $\sin^2 2\theta_{1}, \sin^2 2\theta_2 
\ll 1$, and a strong enhancement of transition probability is due to the 
specific shape of the matter density distribution. Let us consider the three 
layer case  (neutrino oscillations in the earth) with densities 
$N_1$ -  $N_2$ - $N_1$ ($N_1 < N_2$) and concentrate on peaks of the transition 
probability with $P_{max} < 1$ relevant for solar neutrinos. Suppose that 
the neutrino energy is between the MSW resonance energies corresponding to the 
densities $N_1$ and $N_2$ which means that $2 \theta_1 < \pi/2$ and $2 
\theta_2 > \pi/2$. In this case $\sin^2 2\theta_{1,2}\ll 1$ implies that 
$2\theta_1$ is small and $2\theta_2$ is close to $\pi$. If $4\theta_1 + (\pi -
2\theta_2) < \pi/2$\footnote{This condition is equivalent to $\cos(2\theta_2 - 
4\theta_1)< 0$ \cite{PE,Akh2}.} 
the maximal enhancement of the transition probability takes place for the 
values of the oscillation phases $2\phi_i = \pi + 2 \pi k_i$, {\it i.e.} 
for the realization (\ref{res2}) of the parametric resonance condition 
(\ref{res1}).  In this case the transition probability  
given in (\ref{probmax}) can be significantly larger than that in
one layer with a matter of constant density with largest of the two
$\sin^2 2\theta_{i}$ \cite{PE,Akh2}.  

If neutrino energy is above the MSW resonance energies, which means $2 
\theta_1, 2 \theta_2 > \pi/2$, the smallness of $\sin^2 2\theta_{1,2}$ (even 
for large or maximal vacuum mixing) is due to the matter suppression effects.  
Again for $2(\pi - 2\theta_1) - (\pi - 2\theta_2) < \pi/2$ the maximal 
enhancement of the transition probability corresponds to the realization 
(\ref{res2}) of the parametric resonance with probability given in 
(\ref{probmax}) \cite{LS}. 

Notice that for neutrinos traversing the earth the phases $2\phi_i$ are not 
arbitrary and the condition $2\phi_1, 2\phi_2 = (\mbox{odd integer})\times \pi$ 
can be satisfied only approximately. For $2\phi_1, 2\phi_2  \neq (\mbox{odd 
integer})\times \pi$ the transition probability is smaller than (\ref{probmax}). 
In this case, for $\sin^2 2\theta_0<0.03$, the maximum of $P$ is achieved 
for relatively small but non-vanishing values of $X_3$, which corresponds 
to the parametric oscillations with non-maximal depth. 

(ii) For vacuum mixing close to the maximal one, $\sin^2 2\theta_0\aprge 0.9$, 
the MSW resonances in the core and mantle are very wide and therefore 
the mixing angles in medium in the resonance energy interval are also  large: 
$\sin^2 2\theta_1  \sim  \sin^2 2\theta_2 \sim  0.9 - 1$. The change of the 
mixing angle in passing from the mantle to the core or vice versa is small 
and one can consider the earth matter as a single layer with a density close 
to the MSW resonance one. The effect of the matter density profile 
on the transition probability  is small, and what matters is the total 
oscillation phase acquired when neutrinos traverse the earth. The complete 
conversion requires this phase to be an odd integer of $\pi$. In particular, 
for three layers this implies $2(2\phi_1+\phi_2) = \pi(2k+1)$. 
Indeed, large $\sin^2 2\theta_0$ solutions found in \cite{CP2} satisfy this 
equality with a high precision. 

(iii) There are several peaks of transition probabilities with $P(\nu_a\to
\nu_b)=1$ which correspond to intermediate values of the vacuum mixing 
angle, $\sin^2 2\theta_0\simeq 0.15 - 0.6$ \cite{CP2}. These peaks  
are due to an interplay of the effects of large-amplitude oscillations and
specific matter density profile. However, none of the known neutrino anomalies 
can be explained through neutrino oscillations with mixing angles in 
this range. The oscillation solutions of the solar neutrino problem require the 
vacuum mixing angle to be either very small or close to the maximal one; 
the dominant mode of the atmospheric neutrino oscillations requires 
maximal or almost maximal mixing. The mixing angle $\theta_{13}$ governing the 
subdominant $\nu_e\leftrightarrow \nu_\mu$ and $\nu_e\leftrightarrow \nu_\tau$ 
oscillations of atmospheric neutrinos is severely restricted by the CHOOZ 
experiment \cite{CHOOZ} and the solar and atmospheric neutrino observations. 
Nevertheless, the solution with $\sin^2 2\theta_0 \simeq 0.15$ \cite{CP2},
though on the verge of being ruled out by CHOOZ for the range 
of $\Delta m^2$ 
allowed by the Super-Kamiokande atmospheric neutrino data, is at present not 
excluded. It can lead to a significant up-down asymmetry of the e-like 
events in the Super-Kamiokande atmospheric neutrino data (see fig. 7 in
\cite{ADLS}). 

We have shown that the effects discussed in \cite{CP1} are those of the {\it
parametric enhancement} of neutrino oscillations, contrary to the claim of the 
authors that they have found completely new effects which have nothing to do 
with the parametric resonance. Written in the form (\ref{res3}) or (\ref{Z1}) 
the conditions for total neutrino conversion have a clear physical meaning of 
the conditions of the parametric resonance and ($\pi/2+\pi k$) phase of the 
parametric oscillations. 

This work was supported in part by the TMR network grant ERBFMRX-CT960090 of 
the European Union. The work of E.A. was supported by Funda\c{c}\~ao para 
a Ci\^encia e a Tecnologia through the grant PRAXIS XXI/BCC/16414/98.

\end{document}